\documentclass[aps,prx,twocolumn,showpacs,floatfix]{revtex4-2}

\usepackage{graphics,amssymb,amsmath,epsfig,color,textgreek}
\usepackage{graphicx}

\usepackage[dvipsnames]{xcolor}
\usepackage{dcolumn}
\usepackage{bm}
\usepackage[colorlinks=true,citecolor=cyan,breaklinks=true,pdftex]{hyperref}
\usepackage{breakurl}
\hypersetup{colorlinks=true,citecolor=cyan,linkcolor=red,urlcolor=magenta}
\usepackage{braket}

\usepackage[caption=false]{subfig}

\usepackage[dvipsnames]{xcolor}

\newcommand{\ham}{\mathcal{H}}

\DeclareUnicodeCharacter{2212}{-}

\usepackage{comment}

\begin{document}

\title{Subspace Diagonalization on Quantum Computers using Eigenvector Continuation}

\author{Akhil Francis}
\affiliation{Department of Physics, North Carolina State University, Raleigh, North Carolina 27695, USA}

\author{Anjali A. Agrawal}
\affiliation{Department of Physics, North Carolina State University, Raleigh, North Carolina 27695, USA}

\author{Jack H. Howard}
\affiliation{Department of Physics, North Carolina State University, Raleigh, North Carolina 27695, USA}

\author{Efekan K\"okc\"u}
\affiliation{Department of Physics, North Carolina State University, Raleigh, North Carolina 27695, USA}

\author{A.~F.~Kemper}
\email{akemper@ncsu.edu}
\affiliation{Department of Physics, North Carolina State University, Raleigh, North Carolina 27695, USA}

\date{\today{}}
\newcommand{\cmnt}[1]{}

\begin{abstract}
Quantum subspace diagonalization (QSD) methods are  quantum-classical hybrid methods, commonly used to find ground and excited state energies by projecting the Hamiltonian to a smaller subspace. In applying these, the choice of subspace basis is
critical from the perspectives of basis completeness and
efficiency of implementation on quantum computers. 
In this work, we present Eigenvector Continuation (EC) \cite{frame2018eigenvector} as a 
QSD method, where low-energy states of the Hamiltonian at
different points in parameter space are chosen as the subspace
basis. This unique choice enables rapid evaluation of 
low-energy spectra, including ground and nearby excited states,
with minimal hardware effort. As a particular advantage,
EC is able to capture the spectrum across ground state
crossovers corresponding to different symmetry sectors
of the problem. We demonstrate this method for interacting spin models and molecules.

\end{abstract}

\maketitle
\section{Introduction}
Interacting quantum systems, which exhibit some of the
most interesting physics, also pose great challenges. Exactly simulating these systems is generally not feasible using classical computers because of the exponential growth of the Hilbert space with system size. Quantum computers may offer a pathway around the problem  \citep{georgescu2014quantum}, and
a variety of algorithms for obtaining ground and excited states (such as the Variational Quantum Eigensolver (VQE))
\citep{farhi2016quantum,higgott2019variational,kandala2017hardware,peruzzo2014variational,pagano2020quantum} and simulating time evolution
\citep{berry2015simulating,childs2021theory,kokcu2022fixed} have been developed and demonstrated. However, when implemented directly they face difficulties with optimization landscapes \citep{cerezo2021cost,wang2021noise,bittel2021training} and excessive circuit depth that is not suitable for today's quantum hardware, demonstrating a need for more refined algorithms.

Classically, one way to reduce the size of the Hilbert space is by selecting a smaller number of basis states to form a subspace, and projecting the Hamiltonian into that subspace. A typical method is the Krylov subspace method, where the Hamiltonian $\ham$ is expanded in a subspace formed by the repeated action of the $\ham$ on 
a vector \cite{saad1992analysis,hochbruck1997krylov}.
After the subspace projection, the (generalized) eigenvalue problem for this smaller Hamiltonian is solved \citep {lanczos1950iteration, saad1992analysis, frame2018eigenvector, konig2020eigenvector, quarteroni2015reduced} with the help of an overlap or Gram matrix. 

Subspace diagonalization methods can also be implemented efficiently as quantum-classical hybrid algorithms, which we refer to as quantum subspace diagonalization (QSD) methods. In QSD, a quantum computer is used to compute the subspace Hamiltonian and overlap matrix. The quantum computer
thus handles the computations in the larger Hilbert space, and a classical computer can subsequently be used to solve the smaller (generalized) eigenvalue problem. 

Many varieties of QSD methods have already been proposed and implemented \citep{cortes2022quantum,huggins2020non,klymko2021real,stair2020multireference,colless2018computation,bharti2021iterative,McClean:2017,lim2021fast,parrish2019quantum,seki2021quantum,bespalova2021hamiltonian,shen2022real,cortes2022fast}. They vary in terms of the subspace basis selection and implementation scheme. Several of them select basis states by doing real time evolution of a reference state, such as Quantum Krylov subspace methods \citep{cortes2022quantum,stair2020multireference,cortes2022fast}, Variational Quantum Phase Estimation \citep{klymko2021real}, Quantum Power method  \citep{seki2021quantum} and Quantum Filter Diagonalization (QFD) \citep{parrish2019quantum,bespalova2021hamiltonian}. The basis states could also be generated by applying operators to a reference state, for example using Pauli operators \citep{colless2018computation}, creation and annhilation operators \citep{McClean:2017} or elements of the Hamiltonian \cite{lim2021fast,bharti2021iterative}. The subspace diagonalization problem can also be cast as an optimization problem in a variational way as is done in the subspace search variational quantum eigensolver (SSVQE) \citep{nakanishi2019subspace}. 

Once the low energy subspace has been determined, QSD methods can be used to compute quantities such as ground and excited state energies, or for other purposes such as Hamiltonian evolution; examples of this are the Subspace Variational Quantum Simulator (SVQS) \cite{heya2019subspace}, fixed-state Variational Fast Forwarding (fs-VFF) \cite{gibbs2021long}, and Classical Quantum Fast Forwarding(CQFF) \citep{lim2021fast}. Further use has included obtaining excited states \citep{colless2018computation}, and Green's functions \citep{jamet2022quantum}.

Here, we present a QSD approach called \emph{eigenvector continuation} (EC). The EC method (as introduced by Frame et al. \cite{frame2018eigenvector}) uses the fact that continuation of an eigenvector originated by smooth change in the Hamiltonian can be well approximated by a very low dimensional subspace. In other words, the Hamiltonian is considered as a function of one or more parameters, and the low energy eigenstates of the Hamiltonian at a set of parameters (denoted as training points) are chosen as the subspace basis for obtaining low energy state at a target parameter value. EC was demonstrated for computing the ground state energy in three dimensional Bose-Hubbard model and a neutron lattice model \cite{frame2018eigenvector}, and the authors noted the potential advantage of EC over methods such as perturbation and quantum Monte Carlo. Later, a connection of this method with reduced basis methods \cite{almroth1978automatic,quarteroni2015reduced,hesthaven2016certified,herbst2022surrogate} was recognized \cite{bonilla2022training,melendez2022model}. EC has since seen a number of uses in the context of nuclear physics \cite{frame2018eigenvector,konig2020eigenvector,yapa2022volume,demol2020improved}. This method is comparable with other methods in terms of performance \cite{demol2020improved} and it was shown to be extensible to changes in the Hamiltonian
along multiple directions \cite{sarkar2021convergence}. 

The key point is that in EC we select physics-informed subspace basis states, i.e. the low-energy states (training states) at a number of training points in parameter space. These may be obtained analytically if possible, or obtained on the quantum computer using algorithms such as VQE, Quantum Approximate Optimization Algorithm (QAOA), or adiabatic state preparation. Once the subspace basis states at the training points are known, the Hamiltonian (for some parameters not in the training set) and overlap matrices may be measured directly. As we will show below, it is then possible to use these relatively few training points to capture the low-energy spectrum for a wide range of parameters at the cost of a small classical calculation.

One of the main advantages of EC over competing
subspace methods is its ability to
naturally capture orthogonal sectors of the Hilbert
space corresponding to different parameter ranges.
This situation arises, for example, in spin systems
where the desired parameter range spans several 
magnetization sectors, or in quantum chemistry
where the ground state transitions from one
symmetry sector to another (the latter goes by
the general term \emph{conical intersections}).
As we will show, by selecting subspace basis states from
each of the relevant sectors, the method can naturally
span the full parameter range. 

Below, we demonstrate the implementation of EC as a QSD method for spin models from condensed matter (the XY and XXZ models), and for the H$_2$ molecule as a quantum chemistry example. For each case, we select a few training points and demonstrate that the ground state energy or low-energy spectrum can be obtained. For the XY model, we implemented EC on IBM Quantum Computers, and further show that a desired state not in the subspace basis but in the actual Hilbert space can be prepared using Linear Combination of Unitaries (LCU).
An implementation of our algorithm is available \cite{jack_howard_2022_7067454} at \href{https://github.com/kemperlab/EigenvectorContinuation}{https://github.com/kemperlab/EigenvectorContinuation}.


\section{Formalism}

We begin by reviewing the subspace diagonalization method. The issue at hand is solving the time-independent Schr\"odinger equation for a problem with a large Hilbert space dimension ($K$),
\begin{equation}
    \ham \ket{\Psi} = E \ket{\Psi}
    \label{eigen eq}
\end{equation}
Numerically this can be cast as an eigenvalue problem after representing the Hamiltonian as a matrix in an orthonormal basis, or as a generalized eigenvalue problem if the basis is not orthogonal. 
However, given the exponential growth of $K$ with the system size, exact diagonalization on a classical computer becomes prohibitively expensive. On the other hand, if we can find  $M \ll K$ ``basis'' vectors which span the states of interest (target states) then we can project the Hamiltonian to the subspace spanned by the vectors, thus reducing the problem size to something feasible for classical computers.

 Denoting the subspace basis states as  \{$\ket{\phi_p}_{p=1}^M$\}, the time independent Schr\"{o}dinger equation can thus be represented as 
 a generalized eigenvalue equation in this subspace (of dimension $M$)

\begin{align}
    H\ket{\psi} = E \mathcal{S} \ket{\psi},
    \label{schrodinger Eqn}
\end{align}

as long as the state $\ket{\Psi}$ lies within this subspace. Here, $\ket{\psi}$ is the generalized eigenvector in the subspace.
The Hamiltonian matrix $H$ and the overlap matrix (or Gram matrix) $\mathcal{S}$ are
\begin{subequations}
\begin{align}
    H_{ij} &= \Braket{ \phi_i | \ham | \phi_j} 
    \label{eq:hammat}
    \\
     \mathcal{S}_{ij} &= \Braket{ \phi_i | \phi_j}.
    \label{eq:lambdamat}
\end{align}
\end{subequations}

Determining $H$ and $\mathcal{S}$ requires computations in the larger Hilbert space (of dimension $K$). Here, we will perform this calculation using quantum computers.
Following that, the smaller generalized eigenvalue equation (of dimension $M$) is solved in a classical computer.

\begin{figure}[htpb]
\includegraphics[width=0.49\textwidth]{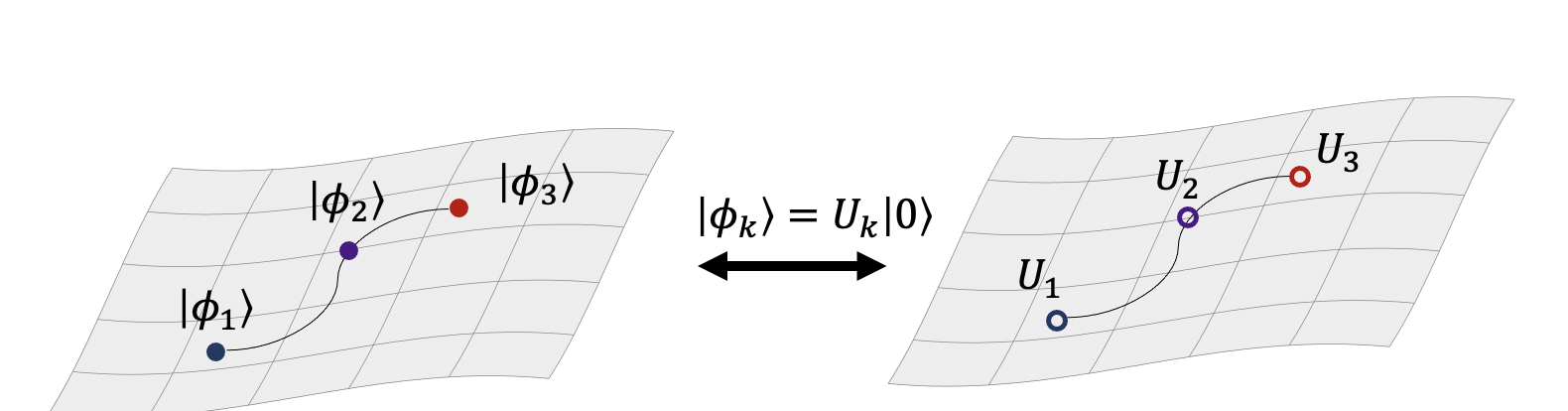}
    \caption{Schematic plot of the EC method. Along the curve in parameter space, $\ket{\phi_3}$ may lie in the span of $\ket{\phi_{1}}$ and $\ket{\phi_2}$. This carries over to the space of unitaries that produce the $\ket{\phi}$ vectors on a quantum computer. States along nearby energy curves could also be taken as training states.}
    \label{fig:schematic evc}
\end{figure}
%
A key aspect of this approach lies in selecting the subspace basis states (\{$\ket{\phi_p}$\}), and different QSD methods select basis differently. Here, we select \{$\ket{\phi_p}$\} based on the EC idea. Consider our Hamiltonian $\ham$ as a function of a parameter $g$, where we want to find the low energy spectrum 
at different values of $g$. For the subspace basis states, we will use low energy eigenstates at some specific parameter values (training points) which we refer to as training states (see Fig.~\ref{fig:schematic evc}). EC \citep{frame2018eigenvector} shows that a few of these states can span a good approximation of the subspace for low energy states at different values of $g$ (target points) as long as these points give rise to smooth changes in the
Hamiltonian. In other words, the EC approach provides eigenstates along the energy curve in parameter space, which is a key advantage. 

The number of the training points needed to have a good approximation for the target state may vary depending on the training and target points. This issue can be addressed variationally; more training states can be added until additional ones no longer affect the target spectrum, indicating convergence has been achieved. In addition, there are situations where adding a low lying state that is not the ground state to the subspace basis set is advantageous \cite{frame2018eigenvector}. We will highlight some of these issues in the examples below.

We now outline some details of the algorithm.
The first step is to obtain the training states, which are eigenstates (and often ground states) of the Hamiltonian at the training points. 
These can be obtained in a variety of ways, depending on the problem,
 including from classical computing techniques like exact diagonalization, or some classical approximation methods. Alternatively, the states can be obtained directly from a quantum computer using VQE, QAOA or adiabatic evolution. Other methods of getting into low energy subspaces like the Schrieffer–Wolff transformation (SWT) in quantum computers \citep{zhang2022quantum} might also be used for basis state preparation.

\begin{figure}[htpb]
\centering
    \includegraphics[]{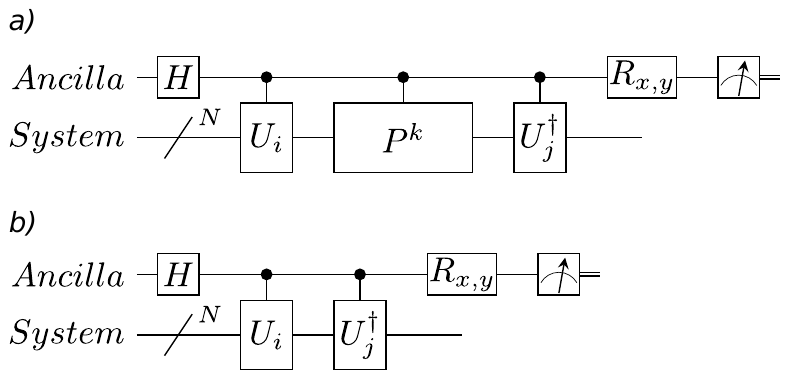}
    \caption{Hadamard test circuits for measuring the elements of the Hamiltonian and the overlap matrix. Here $Ancilla$ denotes the ancilla qubit and $System$ denotes the system qubits. $H$ is the Hadamarad gate and $R_{x/y}$ denotes the rotation gates around X or Y axes to obtain the real and imaginary parts respectively.
    a) Circuit for measuring the Pauli product element $P^k$, of the Hamiltonian (see Eq.~\eqref{Eqn Pauli product}) using the training states $U_{i,j}$ (see Eq.~\eqref{Eqn: training states}) obtain get the projected Hamiltonian (see Eq.~\eqref{eq:hammat}).
    b) Circuit for measuring the overlap component (see Eq.~\eqref{eq:lambdamat})}
\label{fig:measure_circuits}
\end{figure}

The second step of the algorithm for a given target point $g$ in parameter space is to measure the Hamiltonian $H(g)$ (Eq.~\eqref{eq:hammat}) and overlap matrix $\mathcal{S}$ (Eq.~\eqref{eq:lambdamat}) using the subspace basis states. On a quantum computer, the subspace basis states will be represented by unitary operators $U_i$ (see Fig.~\ref{fig:schematic evc}),
\begin{align}
    U_i \ket{0} = \ket{\phi_i },
    \label{Eqn: training states}
\end{align}
and measuring the subspace Hamiltonian and overlap matrices
can be achieved using 
Hadamard test methods (see Fig.~\ref{fig:measure_circuits}).
Measuring the overlap matrix involves only evolution with unitaries which can be directly implemented in a quantum computer. The Hamiltonian is not unitary but they can be written as sums of unitaries. For typical problems of interest, the Hamiltonian on $N$ qubits may be written as
\begin{align}
    \ham(g) = \sum_k^{N_\ham} c_k(g) P^k_N
    \label{Eqn Pauli product}
\end{align}
i.e. a sum of Pauli string operators (elements of the $N$-qubit 
Pauli group $\mathcal{P}_N = \{I,X,Y,Z\}^{\otimes N}$). Typically, as $g$ varies, the Pauli strings remain the same;  only the coefficients change for different target parameters $g$.
Thus, the evaluation of Eq.~\eqref{eq:hammat} can be quite efficient --- each operator in $\ham$ need only be measured once for each pair of subspace basis vectors in the training set.
The changing coefficients only play a role in the classical computation. Similarly, the overlap matrix Eq.~\eqref{eq:lambdamat} needs to be evaluated only once since it does not
depend on $g$.

The Hadamard test for measuring the Hamiltonian elements and the overlap matrix elements using quantum computers can in certain situations be replaced by less costly methods. While Hadamard test methods involve only measuring the ancilla qubit, which reduces the number of shots required, the necessary controlled operations are difficult to implement and significantly increase the depth.
In general, we might not be able to avoid them \cite{araujo2014quantum}, but sometimes reductions are possible \cite{mitarai2019methodology,cortes2022quantum,huggins2020non}. For example, the diagonal components of the Hamiltonian element can be measured with the Hadamard test where control is required only for the unitary Hamiltonian element by preparing the initial state accordingly. 

The third and final step is to solve the generalized eigenvalue equation for the target Hamiltonian, which is done on a classical computer. 
If the subspace is reasonably spanned, and if $H$ and $S$ can be measured accurately, reasonable eigenvalues can be obtained.
However, depending on the choice of training points and the level of machine noise,
the overlap matrix can become ill-conditioned. This can be remedied by altering the choice of training points,
or by treating the matrices to mitigate the noise effects \citep{epperly2021theory}.
Finally, once we have the eigenvector, the wave function in the bigger Hilbert space can be prepared on a quantum computer using Linear Combination of Unitaries (LCU) (see Appendix: \ref{appendix:lcu}).

\section{Results}

To demonstrate the power of EC as a QSD in quantum computers, we demonstrate it in the
context of interacting spin models common in condensed matter systems and often used in quantum
computing --- the one dimensional transverse XY and XXZ models. The former provides an illustrative platform to demonstrate EC. The XXZ model is a spin analogue of the Hubbard model in that it does not admit a free-fermionic solution. Following these, we apply EC to the problem of determining the binding energy of H$_2$ as a function of atomic distance, demonstrating that EC is a valuable technique in this context as well.
%


\subsection{Eigenvector Continuation for the XY model}\label{sec:XY_model}

We consider a one dimensional periodic XY model with nearest neighbor spin interaction, with couplings in the $xy$ plane and with magnetic fields $B_z$ along the $z$-axis and $B_x$ along the $x$-axis. 
The Hamiltonian for this model is

\begin{align}\label{eq:XY_ham}
    \ham_{\mathrm{XY}} &= \sum_{i=1}^{N-1} J\left( X_i  X_{i+1} + Y_i  Y_{i+1} \right) 
    + \sum_{i=1}^{N}B_z  Z_i + \sum_{i=1}^{N}B_x   X_i ,
\end{align}
where $X$, $Y$, $Z$ are the Pauli matrices. Below, we will set $J=-1$, which fixes the energy unit. 

When $B_x=0$, this model can be exactly solved by mapping into a fermionic model via Jordan-Wigner transformation \citep{lieb1961two}. The total magnetization in $z$ direction ($m_z$) commutes with the Hamiltonian, and thus ground state crossovers happen as a function of $B_z$; the magnetization increases with integer units as the energy levels corresponding to different magnetization sectors cross.
For an $N$-site system, there are $N$ the ground state transitions. At $B_z=0$, the ground state of this model has zero magnetization.
With finite $B_x$, the Hamiltonian is no longer
mappable onto a free-fermionic problem, and the ground state crossovers become smooth variations from one magnetization
sector to another.

\begin{figure}[t]
    \centering
    \includegraphics[width=0.49\textwidth]{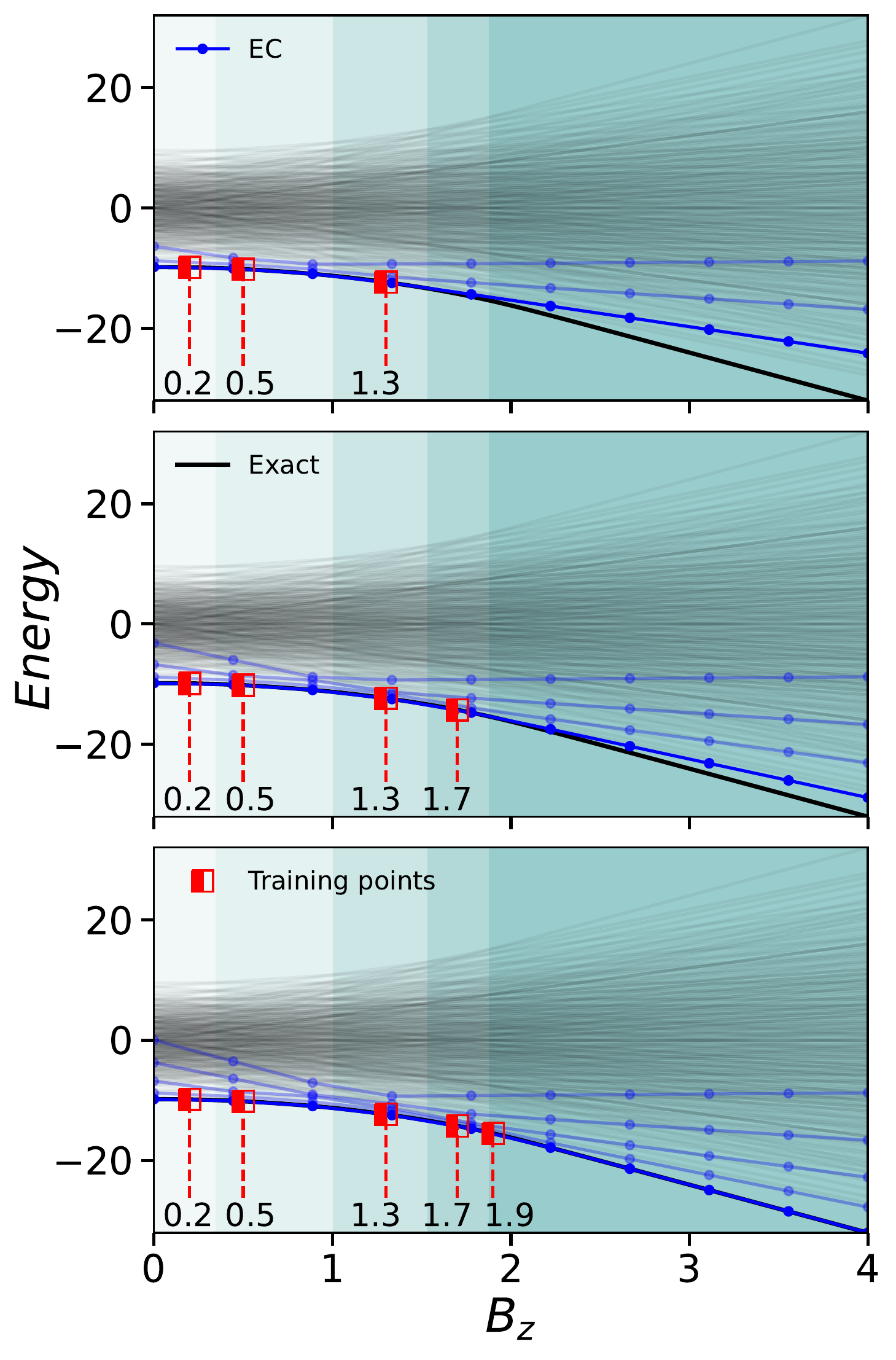}
    
    \caption{Eigenvector continuation for an 8-site XY model with open boundary conditions. Here we have set $B_x=0.1$. Red lines and markers show the training points; the color gradient represents different magnetization sectors separated by level crossings determined with $B_x=0$. The solid black line shows the exact ground state and the partially transparent black lines show the exact excited states. The solid blue line shows the ground state obtained from EC at the target point (blue dots); similarly the partially transparent blue line shows the excited states obtained from EC. }
    \label{fig:3BasesCompletenessN8}
\end{figure}

A typical task would involve finding the ground state as a function of $B_z$, or to locate the transition
between magnetization sectors. Doing so on a quantum computer using VQE or QAOA may involve solving hard optimization problems,
and doing so using adiabatic evolution \citep{richerme2013experimental, francis2022determining}
requires similarly difficult Hamiltonian evolution (in addition to needing to breaking the $z$ magnetization symmetry
with a finite $B_x$).

We will use the ground state at a number of training points as subspace basis states to continue the low-energy spectrum as a function of the parameter $B_z$. Fig.~\ref{fig:3BasesCompletenessN8} demonstrates this for an 8-site XY model. The colored regions in the figure indicate
the different magnetization sectors, and as $B_z$ is increased, the ground state crosses from one sector to the next. To cover the full spectrum, a sufficient number of basis states to span
the eigenstates from each sector is needed;
for this particular problem, choosing the ground states of training points between the level crossings make for a good subspace basis. 
In Fig.~\ref{fig:3BasesCompletenessN8}, we used a finite symmetry breaking field ($B_x=0.1$), but as we will show below in detail, this does not play a critical role in the results.

The figure shows that the ground state is correctly
captured if in the training set has a subspace basis
vector corresponding to the symmetry sector at
the target parameter point, but misses the
ground state if such a vector is not in the basis.
As the number of training points is increased,
the eigenstates from EC follow
the
crossover of the true ground state from one symmetry sector to the next as a function of $B_z$, as well
as a number of excited states.
Note that exact knowledge of the level crossings is not necessary; however, EC does allow us to 
make use of our physical intuition --- in this case, knowing that distinct symmetry sectors exist, even if
we do not know precisely where these start and end.

\subsection{Illustrative 2-site XY model calculation}

\begin{figure}[htpb]
    \centering
     \includegraphics[width=0.49\textwidth,clip=true,
     trim=0 0 0 0]{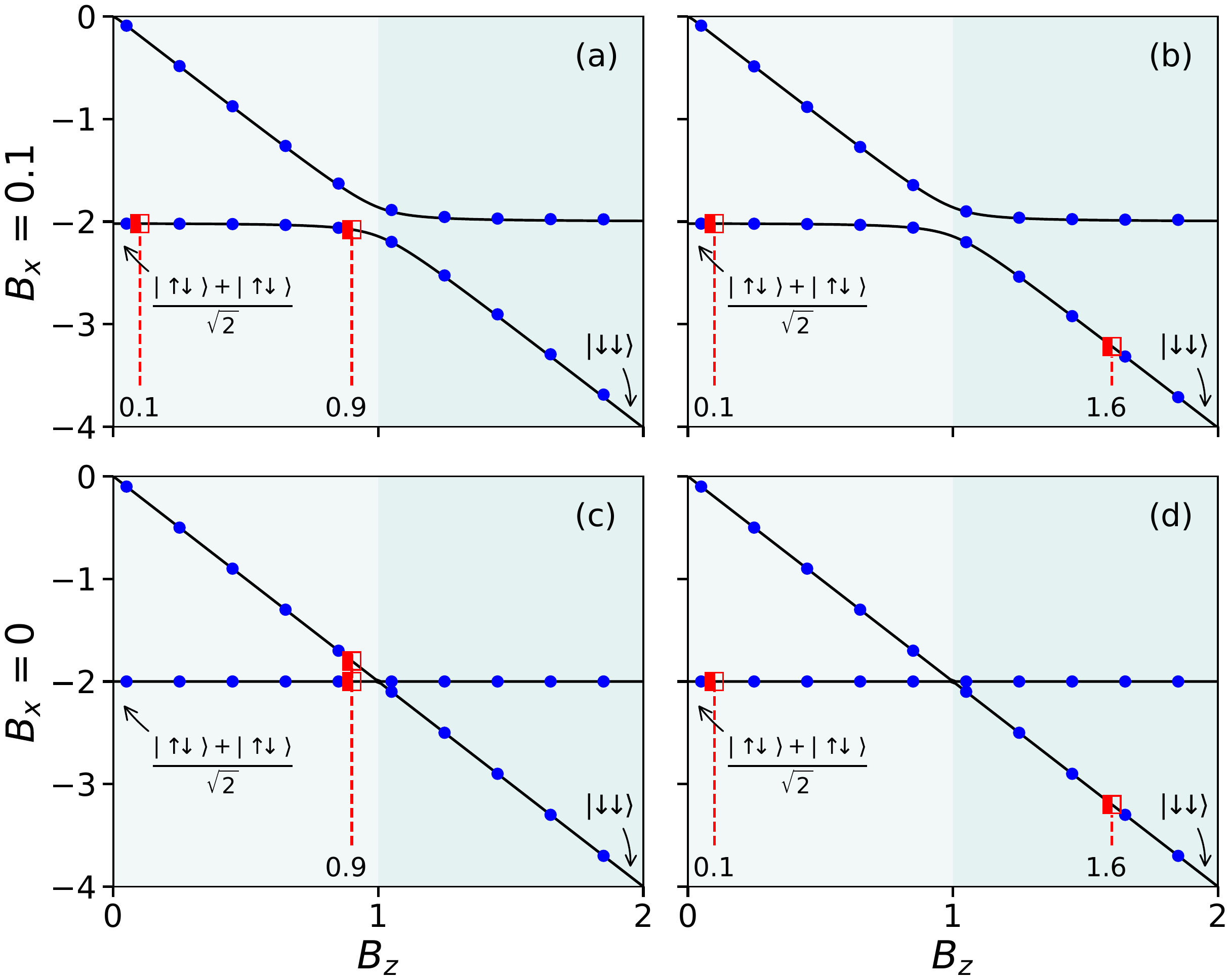}

    \caption{Continuing the eigenspectrum of a two site XY model using two training points (red square and dashed lines). We plot the target eigenvalues (blue solid dots) calculated using EC for four different training sets, with and without in-plane field $B_x$. The exact lowest two eigenvalues are shown using black lines. (a,b) With finite
    $B_x$, training points on either side of the level crossing $B_z=1$ span the subspace. (c,d) When $B_x=0$, training points both sides of the crossing, or using ground and first excited state on one side, are necessary as the basis states to complete the subspace. The states at $B_z=0$ and $B_z=2$ are annotated with their
    $B_x=0$ character.}
    \label{fig:2ExtrapolationInterpolation}
\end{figure}

In order to develop understanding of how EC works, we consider an illustrative example: a two-site XY model system (see Fig.~\ref{fig:2ExtrapolationInterpolation}), which has a single ground state cross over at ($B_z=1$).
We will use two training points to cover the low-energy
spectrum --- in this case, two states of the $S$=1
manifold (the remaining two states are not shown).
Note that this covers half the Hilbert space for
this small example, but as shown in Fig.~\ref{fig:3BasesCompletenessN8}, such a large fraction of coverage is not necessary. We will consider two kinds 
of training sets: (i) both training points on one side of the crossing ($B_z=1$), and (ii) training points on either side of the crossing. 

First, we consider the finite $B_x$ case, with the training points near $B_z=0$ and $B_z=1$ 
[Fig.~\ref{fig:2ExtrapolationInterpolation}(a)].
At $B_z=0.1$,
the training vector (subspace basis vector) is almost entirely in the $m_z=0$ sector; at $B_z=0.9$ the
training vector is a roughly equal mixture of the $m_z=0$ and $m_z=-1$ sectors. These two vectors
are sufficient to span the ground state of entire range of $B_z$ considered here.  Note that the
first excited state is also found as an eigenstate of the subspace Hamiltonian. 
We next choose the training points $B_z=0.1$ and $B_z=1.6$
[Fig.~\ref{fig:2ExtrapolationInterpolation}(b)].
Now, the second vector is mostly in the $m_z=-1$ sector. This set of vectors spans the low-energy subspace equally well.

When $B_x=0$, there is no mixing between the sectors. The ground states at $B_z=0.1$ and $B_z=0.9$ are
identical, and thus insufficient to span the low-energy subspace. This is because in this scenario, ground states on one side have magnetization corresponding to only one symmetry sector and training sets in this symmetry sector cannot span elements in other symmetry sector. Aside from reintroducing the symmetry-breaking
field $B_x$ as above, this can be remedied by using the ground state and the first excited state at one of the training points as subspace basis vectors instead
[Fig.~\ref{fig:2ExtrapolationInterpolation}(c)]. The choice of training points on two sides of the transition,
however, does span the space correctly without further
complications [Fig.~\ref{fig:2ExtrapolationInterpolation}(d)].

\vspace{0.1in}
\subsection{2-site XY model Quantum Computer results}

\begin{figure}[b]
    \centering
   
    \includegraphics[width=0.49\textwidth]{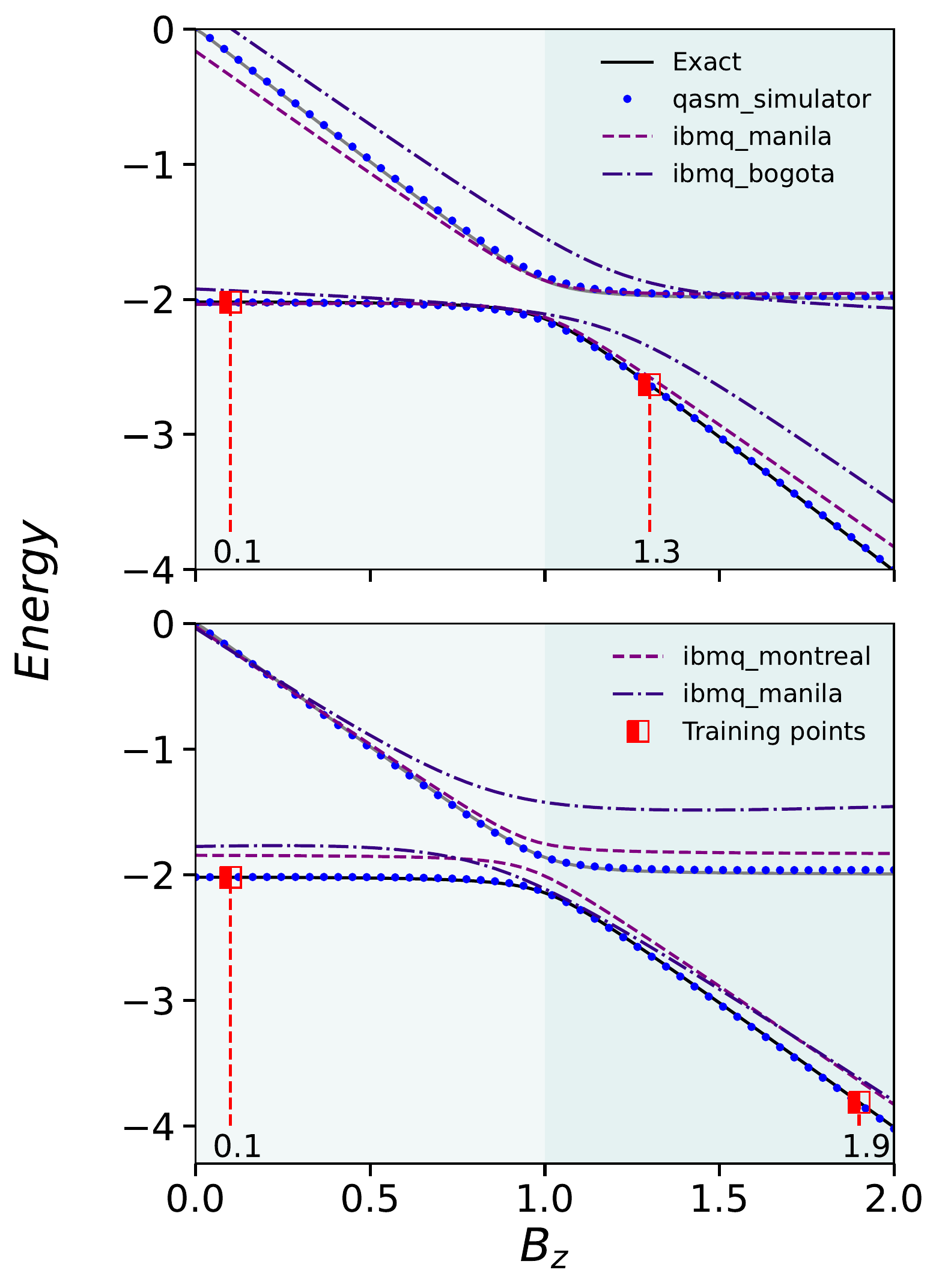}
    \caption{Spectrum of the two-site XY model using eigenvector continuation. Here we have set $J = -1$, $B_x = 0.1$ for all the data points. We have taken two training points $B_z = 0.1, 1.3$, for the top panel and $B_z = 0.1,1.9$ for the bottom panel. Simulator and machine data are taken with 20,000 shots. This model with $B_x = 0$ have a ground state transition happening at $B_z = 1.0$.}
    \label{fig:evc 2site}
\end{figure}
In order to demonstrate the implementation in a NISQ machine we consider a two-site system. We pick two training points and consider target $B_z$ ranging from 0.0 to 2.0. Using these training sets we measure the overlap matrix as well as the Hamiltonian matrix components in Eq.~\eqref{eq:lambdamat}, Eq.~\eqref{eq:hammat}. For Hamiltonian matrix components we measured the Pauli strings separately. Similar to the previous section, we took two sets of two training points near the first cross over (shown Fig.~\ref{fig:evc 2site}). 
We obtained the exact ground states, constructed corresponding unitary operators, and used \emph{search\_compiler} \citep{davis2019heuristics} to find the quantum circuits for measuring the Hamiltonian term and overlap matrices. 
 The data from the quantum computer (see Appendix~\ref{appendix:experimental details}) was corrected with read-out error mitigation \cite{Qiskit-Textbook}. 
Note that since the target Hamiltonians can
 be generated entirely from the individual Hamiltonian term's matrices which are measured only once, no additional quantum computer runs are necessary.

 With the Hamiltonian term and overlap matrices in hand, we
 obtain the spectrum for broad range of
 $B_z$ values. This is shown in Fig.~\ref{fig:evc 2site} using a simulator and three hardware quantum computers ({\em ibmq\_montreal}, {\em ibmq\_bogota} and {\em ibmq\_manila}). The target energies obtained from IBM quantum computers, shown as dashed and dashed-dotted lines, are close to the simulator values. The simulator correctly captures the low-energy spectrum --- the hardware similarly captures the correct behavior, but with some amount of discrepancy most likely associated with the various forms of hardware noise.

\begin{figure}[t]
    \centering
   
    \includegraphics[width=0.49\textwidth]{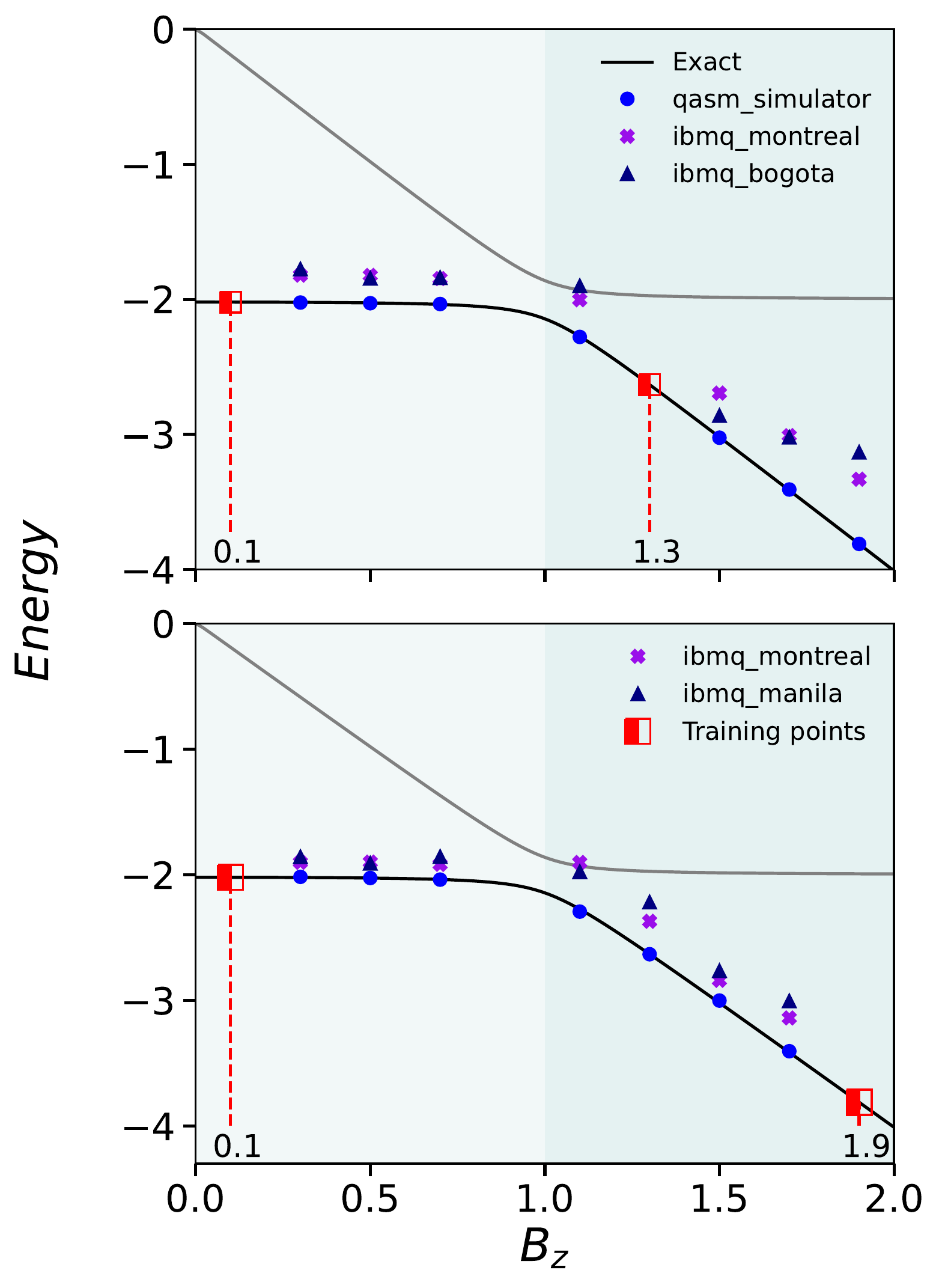}
    \caption{Energy eigenvalues obtained from ground states prepared using LCU based on EC results. Here we have set $J = -1$, $B_x = 0.1$ for all the data points. We have taken two training points $B_z = 0.1, 1.3$, for the top panel and $B_z = 0.1,1.9$ for the bottom panel. Simulator and machine data are taken with 20,000 shots.}
    \label{fig:lcu 2site}
\end{figure}

In addition to obtaining the low-energy spectrum, it is desirable to be able to prepare the obtained eigenstates
on the quantum computer for future use (e.g. to measure its properties). We can prepare the state in a quantum
computer using block encoding using Linear Combination of Unitaries (LCU) \citep{childs2012hamiltonian}. The target ground state can be expressed as a linear combination of the training states (\{$\ket{\phi_p}$\}), with the coefficients obtained from the
diagonalization of the subspace Hamiltonian. This requires a number additional ancilla qubits depending upon the number of states to be added (see Appendix~\ref{appendix:lcu}).

Working again with the 2-site XY model, we use LCU to prepare the ground state on a quantum computer using the eigenvector obtained from the subspace diagonalization on a classical computer. We subsequently measure the energy of the prepared state, and compare this to the exact result in Fig.~\ref{fig:lcu 2site}. The energies obtained from the quantum computer match reasonably well with the simulator values. 

\subsection{XXZ Model}

\begin{figure}[b]
    \centering
    
    \includegraphics[width=0.5\textwidth]{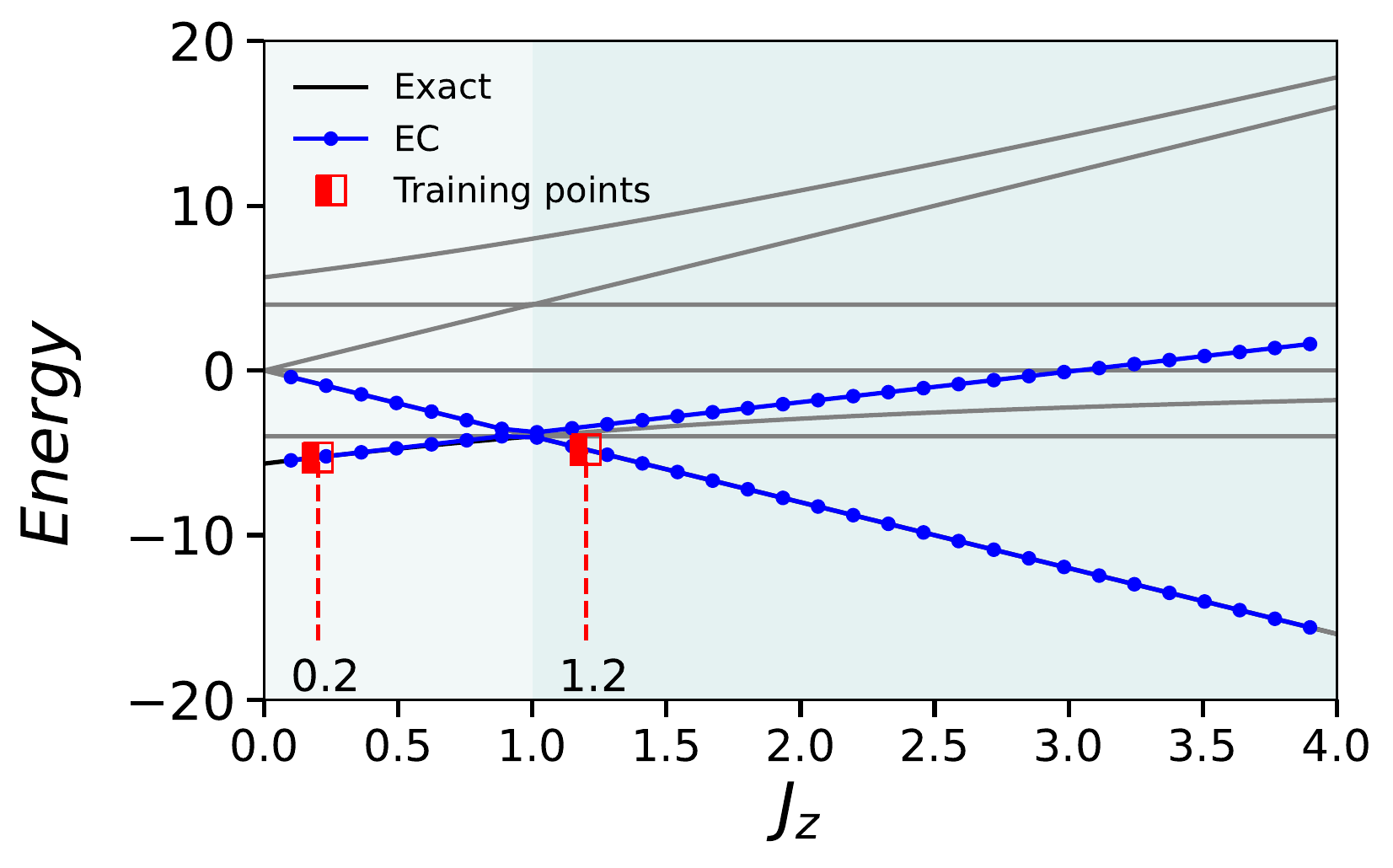}
    \caption{Spectrum of the 4-site XXZ model using eigenvector continuation. We have selected two ground states on both sides of the transition as training points (red square and dashed line). Blue dotted lines represents the target eigenvalues obtained from EC. Black and grey lines represent the exact curves.}
    \label{fig:XXZ}
\end{figure}

EC works equally well for more complex models such as the XXZ model. The Hamiltonian is given as
\begin{align}
    \ham_{\mathrm{XXZ}} &=  \sum_i^{N-1} J \left( X_i  X_{i+1} + Y_i  Y_{i+1} \right)  - J_z   Z_i  Z_{i+1}.
\end{align}

While this is a similar nearest neighbor interacting spin model, here the presence of the $J_z$ term prevents solution by fermionization.
In the fermionic language (after Jordan-Wigner transformation) 
the $J$ term is a nearest neighbor hopping term, while the $J_z$ term is a quadratic term in number operators, mapping the problem onto interacting fermions.

Here, we set $J=1$ (this fixes the energy units) and select $J_z$ as the parameter for EC, starting from the non-interacting limit $J_z=0$. At $J_z = J$ the system undergoes a ground state cross over (see Fig.~\ref{fig:XXZ}); when $J>J_z$ the spins
are oriented in the XY plane with zero $z$ magnetization, which transitions to spins orienting in the $z$ direction when $J_z > J$. As shown in Fig.~\ref{fig:XXZ} EC correctly captures
the ground state spectrum. Here, we have selected the subspace basis states to be ground states on both sides of the transition. Equivalently we could have also selected excited states on one side of the transition, or introduced a symmetry breaking field.


\subsection{Hydrogen Molecule}

Finally, we move on to the quantum chemistry example of a hydrogen molecule (H$_2$). The electronic structure problem of molecules is a well developed topic in quantum chemistry \citep{szabo2012modern}. 
QSD methods are also applied here, such as Quantum Krylov methods \citep{stair2020multireference}, and Quantum Filter Diagonalization methods \citep{parrish2019quantum, cohn2021quantum}. 
Here we demonstrate EC for the hydrogen molecule,
with the interatomic distance as the Hamiltonian parameter to vary.  Thus, we aim to calculate the binding energy curve using only a few select points along the way.

We used {\em Qiskit-nature} for the simulation, which works with the electronic structure Hamiltonian in second-quantized form (see Appendix: \ref{appendix:quantum chemistry}). The geometry is specified by the interatomic distance (which is the parameter $g$). We used the STO-3g which is a basis where each element is composed of three Gaussians; with a larger basis set more accuracy can be achieved in finding the energies, but this set results in a sufficiently accurate Hamiltonian for demonstration purposes \cite{szabo2012modern}.  The Hamiltonian is mapped into the qubit space using a parity mapper and using two-qubit reductions which brings the Hamiltonian to a two qubit problem \cite{Qiskit-Textbook}. 
\begin{figure}[htpb]
    \centering
    \includegraphics[width=0.49\textwidth]{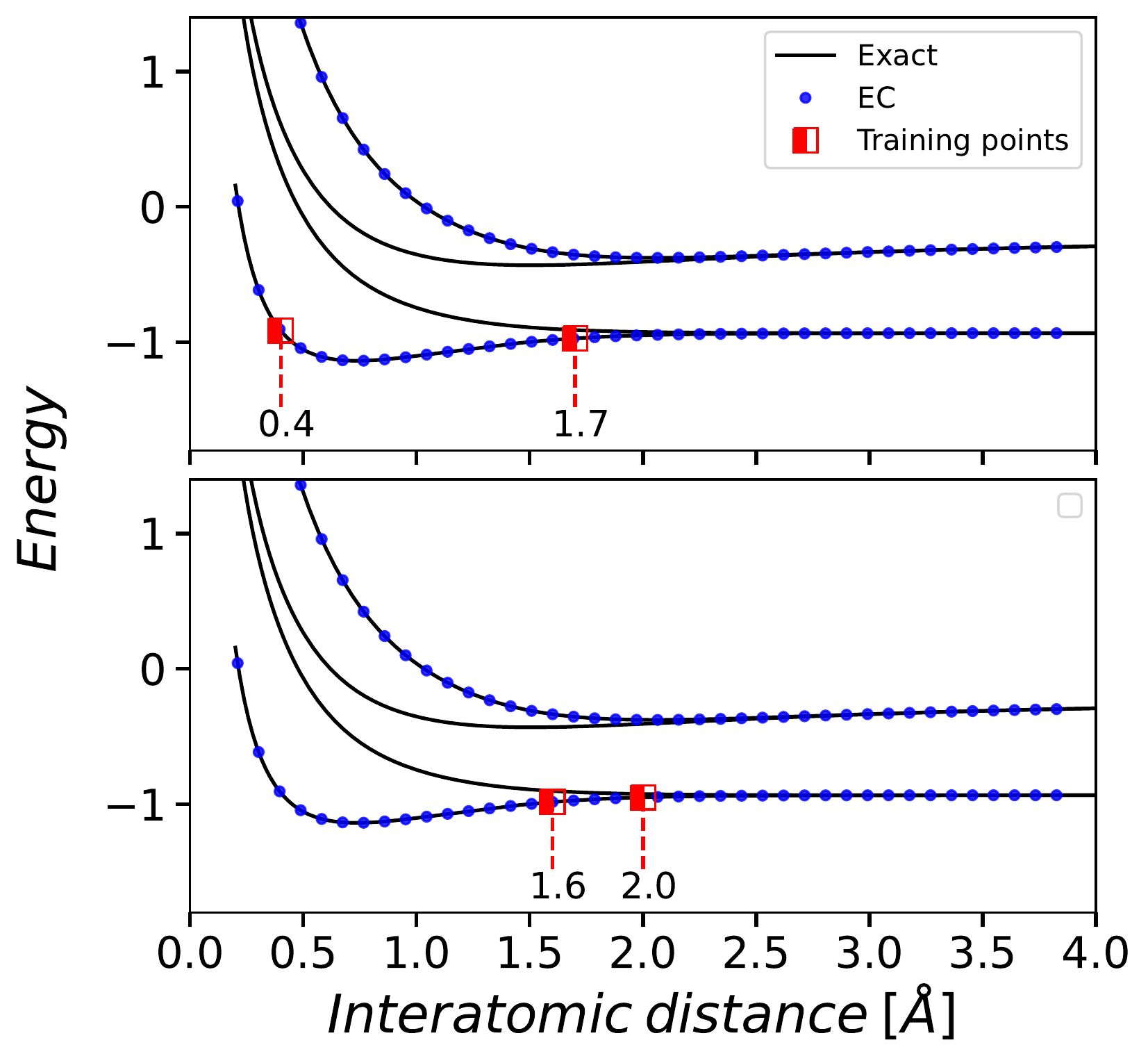}
    \caption{Demonstrating EC method for a hydrogen molecule. Here the electronic Hamiltonian is mapped as a qubit problem from the fermionic degrees of freedom using $Qiskit−nature$. The Hamiltonian parameter is the inter atomic distance and energy is given in units of hartree. We have selected two sets of two training points (red squares) and are able to reproduce the ground state energies using EC. Blue lines represent energies from EC. Black lines represent energies from the exact diagonalization of the two-qubit Hamiltonian along with the nuclear energies.}
    \label{fig:5H2}
\end{figure}

We select two training points and used the ground states as our subspace training vectors for EC (see Fig.~\ref{fig:5H2}). With these points, EC produces a binding energy that, after adding the nuclear energy to the electronic one, matches the exact values when either both training points are on one side of the minimal energy point or on both sides. In this particular case, the H$_2$ Hamiltonian in the minimal basis is represented by a two dimensional subspace composed of the Hartree-Fock (HF) ground state and the doubly excited state \citep{szabo2012modern}. Thus with only two training states we can span the subspace. In general for other molecules such a decoupled structure may not exist;
by Brillouins theorem, Hartree-Fock ground states do not couple directly with single excitations, but they could be indirectly coupled through double or higher excitations \cite{szabo2012modern}. In such a case, more subspace basis states would be needed.

It should be noted that, unlike in the spin model case,
our demonstration naively uses different computational basis states at different training points. This is a particular
issue that can arise in EC because the representation of
the Hamiltonian can change as the parameter of choice is
varied.
Here, it arises because the Fock operators now depend on the molecular geometry, and converting to the spin basis will naturally lead to a computational basis set that depends upon the molecular geometry for the different training points. Nevertheless we obtain the correct results because the eigenvalues are independent of the choice of the basis set within the training subspace as long as we faithfully represent the target state (see Appendix~\ref{appendix:basis inconsistancy} for further details).
Put differently, one can think of being ignorant about the basis representation as equivalent to selecting a different training state (this may not be a low lying eigenstate, but one that represents the same training subspace). This will give the same eigenvalues as long as we are able to represent the training subspace faithfully.
However, if this approach is to be continued and we attempt to construct the obtained states using LCU
as was done with the XY model, then the basis set consistency needs to be handled carefully.
%

\section{Discussion}

We have introduced eigenvector continuation (EC) as a method
for performing quantum subspace diagonalization (QSD).
Using this method we can continue along the low energy curves as a function of a Hamiltonian parameter using the low energy eigenstates at few parameter values. Selecting the subspace using these physics informed basis states is the key difference from other QSDs.

As with any subspace method, it has a wide range of potential
use cases from across different domains. In particular, the
method can be quite useful in situations where the physics
or chemistry dictates a rapid change in eigenvector character.
This was captured in our demonstration case of the XY model,
where a rapid change in magnetization occurs; another similar
area could be found in the study of conical intersections
in chemistry \citep{matsika2007conical}.

In order to demonstrate our QSD approach, we have used a circuit compiler to construct the gates from the unitary matrix representation for our small size system. Such classical compilers were employed only for the ease of demonstration, and will not scale to large systems. Hence, moving forward, we need to use quantum methods like adiabatic time evolution or quantum classical hybrid methods like VQE or QAOA to get the training states and their quantum gates representation. Classical methods could also be employed to get the training states which could then be converted to quantum gates.  Once the circuit representations for the training states are obtained, these can be used directly in the LCU circuits to produce the desired state on the quantum computer.

One of the benefits of QSDs is that they can tolerate some errors in basis set as long as the the subspace can be
spanned \citep{McClean:2017}; this can be advantageous as usually VQE or similar approaches yield approximate ground states. 
For accurate results, the Hamiltonian and overlap matrices need to be measured accurately \citep{huggins2020non}. However, 
shot noise and other errors pose limitations on doing so, 
and this may lead to problems with ill-conditioning of the overlap matrix. This might be an issue for the generalized eigenvalue problem, but empirically QSDs appear to work
directly or by mitigating some of the noise via thresholding
\citep{epperly2021theory}.
We leave the investigation of the noise-resistance of EC
for future work.

Finally, 
because of the limitations of current level NISQ machines,we have analysed and demonstrated EC only in small systems. 
However, the hardware is rapidly advancing, and it is
prudent to keep an eye towards the future capabilities.
As the number of qubits grows larger, there will be a need for
methods that scale to such a size. QSD methods may play a role
here --- and in particular, a QSD method such as EC that can
form eigenstates for interacting systems out of states that are
relatively easy to prepare can prove to be beneficial.

%

\section*{Acknowledgements}
We acknowledge helpful discussions with Roel van Beeumen, Daan Camps, Thomas Steckmann, Daniel Claudino, Yan Wang, and Carlos Mejuto Zaera. This work was supported by the National Science Foundation under grant no. NSF DMR-1752713. We acknowledge the use of IBMQ via the IBM Q Hub at NC State for this work \citep{Qiskit_full}. The views expressed are those of the authors and do not reflect the official policy or position of the IBM Q Hub at NC State, IBM or the IBM Q team. 

\bibliography{ref,qst}
\bibliographystyle{apsrev4-2}

\clearpage

\appendix

\renewcommand\thefigure{A\arabic{figure}}  
\setcounter{figure}{0}

\section{Linear Combination of Unitaries (LCU)}
\label{appendix:lcu}
The details of LCU are described in  \citep{childs2012hamiltonian}; we include a brief overview here 
for completeness.
Given two unitaries ($U_a$ and $U_b$), we can construct the following LCU using an ancilla qubit,
\begin{align}
    \frac{k}{k+1} U_{a}+\frac{1}{k+1} U_{b},
\end{align}
where $k\geq 0$.
\begin{figure}[htpb]
    \centering
    \includegraphics{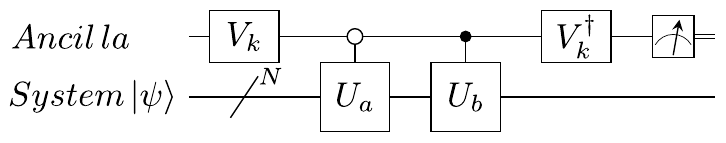}
\caption{LCU circuit for two unitaries. After selecting on the $|0\rangle$ state in the ancilla qubit, we have the state proportional to $k U_a +U_b$ in the $N$ system qubits. }
\label{fig:LCU circuit}
\end{figure}
In order to achieve this we use the usual LCU quantum circuit (see Fig.~\ref{fig:LCU circuit}), where
the resulting unitary is applied to the state $\ket{\psi}$. The ancilla qubit is initialized with the unitary $V_{\kappa}$ 
\begin{align}
V_{k}:=\left(\begin{array}{ll}
\sqrt{\frac{k}{k+1}} & \frac{-1}{\sqrt{k+1}} \\
\frac{1}{\sqrt{k+1}} & \sqrt{\frac{k}{k+1}}
\end{array}\right) .
\end{align}

The LCU circuit just before measurement transforms the initial state as,

\begin{align*}
|0\rangle \otimes |\psi\rangle \mapsto |0\rangle \otimes \left(\frac{k}{k+1} U_{a}+\frac{1}{k+1} U_{b}\right)|\psi\rangle + \\ |1\rangle \otimes \frac{\sqrt{k}}{k+1}\left(U_{b}-U_{a}\right)|\psi\rangle.
\end{align*}
By selecting only the $|0\rangle$ state for the ancilla qubit we have the desired state in the system qubits, and we use this to prepare the ground state for the two site system. We have two basis states $|\psi_a\rangle$ and $|\psi_b\rangle$ that can be prepared by applying $U_0$ and $U_1$ respectively to the initial $|0\rangle$ state on the system qubits. After diagonalizing the subspace Hamiltonian we have the ground state given as $(r_0\exp{(i\theta_0)},r_1\exp{(i\theta_1)})^T$, where the coefficients are complex. We can prepare the ground state using LCU, by absorbing the phases of the coefficients into the basis unitaries such that $U_a = \exp{(i\theta_0)} U_0$ , $U_b = \exp{(i\theta_1)} U_1$ and setting $k = \frac{r_0}{r_1}$, and $\ket{\psi} = \ket{0}$. When we have more unitaries to combine this is dealt by introducing more ancilla qubits. 

\section{Quantum Chemistry in Qiskit}
\label{appendix:quantum chemistry}

The molecular Hamiltonian is usually separated into electronic and nuclear part separately after using the Born - Oppenheimer approximation as nuclear mass is much larger than that of electrons \citep{szabo2012modern}. Then a  set of basis states are selected upon which electronic Hamiltonian can be expressed in the second quantized form after selecting a basis set $\{\chi_{p}\}$ as \citep{whitfield2011simulation} 
\begin{align}\label{eq:H_2_fermionic}
H=\sum_{p, q} h_{p q} a_{p}^{\dagger} a_{q}+\frac{1}{2} \sum_{p, q, r, s} h_{p q r s} a_{p}^{\dagger} a_{q}^{\dagger} a_{r} a_{s},
\end{align}
where $a^{\dagger}$, $a$ are the creation and annihilation operators, and $p,q,r,s$ are the general indices of operator basis set.
The coefficient of the quadratic term in the Hamiltonian comes from single electron integral that accounts for the  kinetic energy part and the electron-nuclear attraction part of the Hamiltonian as given below:
$$
h_{p q} = \int \mathrm{d} \mathbf{x} \chi_{p}^{*}(\mathbf{x})\left(-\frac{1}{2} \nabla^{2}-\sum_{\alpha} \frac{Z_{\alpha}}{r_{\alpha, \mathbf{x}}}\right) \chi_{q}(\mathbf{x})
$$
The coefficient of the quartic term comes from the two-electron integral to account for the electron -electron repulsion part.
$$
h_{p q r s} = \int \mathrm{d} \mathbf{x}_{1} \mathrm{~d} \mathbf{x}_{2} \frac{\chi_{p}^{*}\left(\mathbf{x}_{1}\right) \chi_{q}^{*}\left(\mathbf{x}_{2}\right) \chi_{r}\left(\mathbf{x}_{2}\right) \chi_{s}\left(\mathbf{x}_{1}\right)}{r_{12}}
$$

Once we specify the molecular geometry and the input basis set, solving the Hartree-Fock equations iteratively yields canonical orbitals, where the lowest energy orbitals are occupied, which is filled according to the number of electrons occupied. Filling the lowest states gives the Hartree-Fock ground state while excitations from these states represent the next possible states. The configuration Interaction Hamiltonian is written in this subspace, which is solved to get the best possible ground state. 
All these calculations can be done using the $\mathrm{PYSCF}$ \cite{sun2020recent} package which has been incorporated into the $\mathrm{Qiskit-nature}$ package. We start by specifying the molecular geometry and the input basis which used by the $\mathrm{PYSCF}$ driver in $\mathrm{Qiskit-nature}$ which then maps to the qubit subspace once we specify the type of mapping and denote symmetry be used to reduce the problem size. For hydrogen atom we selected STO-3g basis set and used the in-built parity mapper and two-qubit reductions to map to a two qubit problem as given below,

\begin{align}\label{eq:h2_hamiltonian}
\begin{split}
    H(R) & = a_1(R) II + a_2(R) ZI + a_3(R) IZ
    \\ &+ a_4(R) ZZ + a_5(R) XX .
\end{split}
\end{align},

where $a_1,a_2,a_3,a_4,a_5 \in \mathbb{C} $ and $R$ is the interatomic distance.


\section{Basis consistency}
 \label{appendix:basis inconsistancy}
 One peculiar aspect of implementation of eigenvector continuation is that the underlying basis of the various training points may be inconsistent. To illustrate this, let us introduce the following notation: $g$ as a parameter that the Hamiltonian depends on, $\ham(g)$ as the Hamiltonian of the system. Further, define $\ket{\Psi_i(g)}$ as the $i$th eigenket of $\ham(g)$, and $\ket{n,g}$ as the state represented via the $n$th computational basis ket with the corresponding $g$ value.
For the XY model given in \ref{sec:XY_model}, the parameter is $g = B_z$, and $\ham(g)$ is given in \eqref{eq:XY_ham}. For the H$_2$ molecule, $g =R$ which is the distance between the two hydrogen nuclei, and $\ham(g)$ is given at Eq.~\eqref{eq:h2_hamiltonian}.

The difference between these models lies in what computational basis states represent. For the XY model, $\ket{000}$ represent $\ket{\uparrow \uparrow \uparrow}$ for any value of $g$. Therefore we have, for any computational basis ket, $\ket{n,g_0} \equiv \ket{n,g_1}$ for any $g_0, g_1$ values. In contrast, for the hydrogen molecule, qubit represent Hartree-Fock orbitals centered on the hydrogen nuclei, and computational basis states are representing states in which some of the orbitals are occupied by electrons, and some are not.
Because the orbitals are not the same for different values
of $g$, the states that the computational basis states represent are also not equivalent for different values of $g$, i.e. $\ket{n,g_0} \not\equiv \ket{n,g_1}$. 
This is the case not only for $\text{H}_2$ molecule, but for any molecular Hamiltonian that is expressed in an atom-centered basis. Considering that in EC we take training points for different values of $g$, finding the ground states and using these to span the states for other $g$ values, the inequivalence $\ket{n,g_0} \not\equiv \ket{n,g_1}$ reveals the need for a definition of how to take linear combinations of the computational basis stats for different $g$ values. 

This need actually originates from the fact that the Hilbert spaces for different $g$ values are not actually the exact same space. What we have is not just one Hilbert space, but a set of Hilbert spaces labeled with $g$. For each individual Hilbert space, there is a well defined inner product. However, there is no `unique' definition of inner product between states in different Hilbert spaces with different values of $g$. This gives us a freedom to choose how to `connect' these Hilbert spaces together, i.e. we are allowed to choose the values of $\Braket{n,g_0|n',g_1}$ for $g_0 \neq g_1$. 
This is closely related to the concept of tangent spaces and parallel transport in differential geometry.

This degree of freedom is reminiscent from the unitary transformation that preserves the dynamics:
\begin{align}
    \ham &\to U \ham U^\dagger, &\ket{\Psi} \to U \ket{\Psi}.
\end{align}
Here $\ham$ is the Hamiltonian that is given to EC as input such as \eqref{eq:XY_ham} and \eqref{eq:h2_hamiltonian}.
If we choose a $g$-dependent unitary $U$, we can rotate each Hilbert space freely, and therefore alter the inner product of two states with different labels, without changing the dynamics of any system. This means that we can rotate the training ground states freely with any unitary matrix. 

An interesting implication of this degree of freedom is the following. Say we train our system with $g_1$ and $g_2$ values, and obtain $\ket{\Psi_1}$ and $\ket{\Psi_2}$ as our training states. If we chose $U(g)$ randomly, we could generate any random state instead of $\ket{\Psi_1}$ and $\ket{\Psi_2}$. For most of the choices of $U(g)$, this would lead to poor results with eigenvector continuation, and the algorithm would require more training points. On the other hand, for some choice of $U(g)$, we can conclude that far fewer training points are sufficient to apply eigenvector continuation.

For the hydrogen molecule example above, the naive choice $U(g) = \mathrm{id}$ works with 2 training points. The Hamiltonian structure in Eq.~\eqref{eq:h2_hamiltonian} 
implies that there are two decoupled subspace (1st and 4th basis states, 2nd and 3rd basis states) of dimension two. Since any two non identical training states can fully span the ground state subspace in the qubit space, EC requires only two training points for this case.

In other words, implementing EC for quantum chemistry problems requires handling of the basis inconsistency which naturally arises while using packages like $\mathrm{PYSCF}$ \cite{sun2020recent} originating from the fact that the electronic basis states depends on the Hamiltonian parameter (here, molecular geometry).
If we want to use the ground states (or low energy states) of the electronic problem at training points as the training vectors as EC  does, then we need to appropriately transform these training states to the space represented by target parameter. Alternatively, one may naively treat the training vectors as elements in the Hilbert space of to the target parameter (the $U(g)=\mathrm{id}$ choice), and these might still be good training vectors even if the resulting vectors do not correspond to the low energy eigenstates at the training points, as long as a sufficient space is spanned.

This issue of basis inconsistency does not arise for other QSD methods \cite{klymko2021real,cortes2022quantum,parrish2019quantum, cohn2021quantum} because all basis vectors corresponded to same molecular geometry. This mixing of basis vectors corresponding to different molecular geometries by the EC approach makes this method distinct from other subspace methods.
%


\section{Experimental Details}
\label{appendix:experimental details}

\begin{figure}[!htb]
    \includegraphics[width=0.49\textwidth]{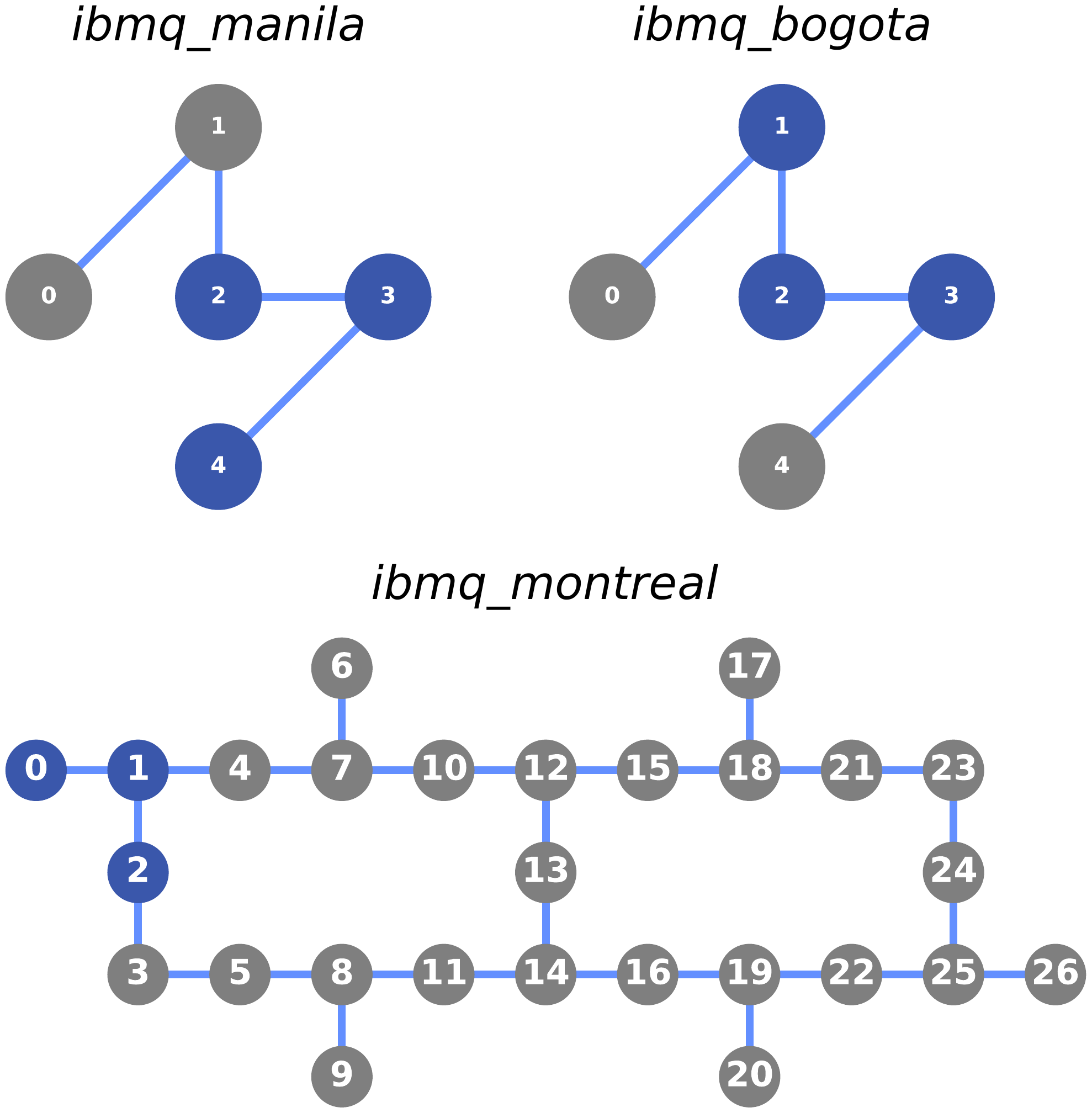}
    \caption{Layouts of $\emph{ibmq\_manila}$, 
    $\emph{ibmq\_bogota}$, and $\emph{ibmq\_montreal}$. For the
    calculations in the main text, qubits $(2,3,4)$ on $\emph{ibmq\_manila}$, $(1,2,3)$ on $\emph{ibmq\_bogota}$, 
    and $(0,1,2)$ on $\emph{ibmq\_montreal}$ were used.}
    \label{fig:backendlayouts}
\end{figure}

All the hardware quantum computer experiments were performed on IBM quantum machines (see Fig.~\ref{fig:backendlayouts}). 
The calibration data for the quantum machines on the dates when experiments are performed along with data for figures are provided in the  \href{https://github.com/kemperlab/EigenvectorContinuation/tree/main/examples/paper\_plots/data\_figs\_calibs}{GitHub repository}. Relevant calibration data for the qubits used are also given in the following Table \ref{table:bogota}, \ref{table: manila}, \ref{table: montreal 1}, \ref{table: montreal 2}, \ref{table: montreal 3}.
\begin{table}[htpb]
\centering
\begin{tabular}{|c| c| c |c| c |c| } 
 \hline
 Qubits & T1 ($\mu s$) & T2 ($\mu s$) & readout &  CNOT & CNOT   \\
  & &  & error & connection & error \\
  \hline \hline
 1 & 105.77 & 45.05 & 0.0725 & 1-2 & 0.006895 \\ 
   \hline
 2 &111.6 & 178.6 & 0.0319 &  2-3 & 0.009015 \\ 
 \hline
 3 &110.1 & 194.67 & 0.0265 & 3-2 & 0.009015 \\ 
 \hline
\end{tabular}
\caption{Calibration data for \emph{ibmq\_bogota} on 25th Feb 2022.}
\label{table:bogota}
\end{table}
\begin{table}[htpb]
\centering
\begin{tabular}{|c| c| c |c| c |c| } 
 \hline
 Qubits & T1 ($\mu s$) & T2 ($\mu s$) & readout &  CNOT & CNOT   \\
  & &  & error & connection & error \\
  \hline \hline
 2 & 118.63 & 22.6 & 0.0368 & 2-3 & 0.006757 \\ 
  \hline
 3 & 182.67 & 65.14 & 0.021 &  3-4 & 0.005721 \\ 
 \hline
 4 & 157.35 & 40.39 & 0.0199 & 4-3 & 0.005721 \\ 
 \hline
\end{tabular}
\caption{Calibration data for \emph{ibmq\_manila} on 25th Feb 2022.}
\label{table: manila}

%

\centering
\begin{tabular}{|c| c| c |c| c |c| } 
 \hline
 Qubits & T1 ($\mu s$) & T2 ($\mu s$) & readout &  CNOT & CNOT   \\
  & &  & error & connection & error \\
  \hline \hline
 0 & 104.03 & 52.11 & 0.0084 & 0-1 & 0.0005327 \\ 
  \hline
 1 & 84.27 & 21.64 & 0.0135 &  1-2 & 0.007531 \\ 
 \hline
 2 & 86.03 & 108.64 & 0.014 & 2-1 & 0.007531 \\ 
 \hline
\end{tabular}
\caption{Calibration data for \emph{ibmq\_montreal} on 25th February 2022.}
\label{table: montreal 1}

%

\centering
\begin{tabular}{|c| c| c |c| c |c| } 
 \hline
 Qubits & T1 ($\mu s$) & T2 ($\mu s$) & readout &  CNOT & CNOT   \\
  & &  & error & connection & error \\
  \hline \hline
 0 & 88.47 & 52.11 & 0.0096 & 0-1 & 0.0005891 \\ 
  \hline
 1 & 82.88 & 21.64 & 0.0171 &  1-2 & 0.007801 \\ 
 \hline
 2 & 81.07 & 109.58 & 0.0122 & 2-1 & 0.007801 \\ 
 \hline
\end{tabular}
\caption{Calibration data for \emph{ibmq\_montreal} on 26th February 2022.}
\label{table: montreal 2}

%

\centering
\begin{tabular}{|c| c| c |c| c |c| } 
 \hline
 Qubits & T1 ($\mu s$) & T2 ($\mu s$) & readout &  CNOT & CNOT   \\
  & &  & error & connection & error \\
  \hline \hline
 0 & 103.11 & 95.41 & 0.0311 & 0-1 & 0.0017 \\ 
  \hline
 1 &137.6 & 43.6 & 0.0153 &  1-2 & 0.01084 \\ 
 \hline
 2 &112.84 & 46.65 & 0.0197 & 2-1 & 0.01084 \\ 
 \hline
\end{tabular}
\caption{Calibration data for \emph{ibmq\_montreal} on 28th March 2022.}
\label{table: montreal 3}
\end{table}

\end{document}